\documentclass{ethpaper}
\usepackage{graphicx}
\usepackage{here}
\usepackage{amssymb}
\usepackage{amsmath}
\usepackage{subfigure}
\usepackage{lineno}

\begin{document}
\begin{titlepage}
\ethnote{}
\title{Predicting hadron-specific damage from fast hadrons in crystals for calorimetry
}
\begin{Authlist}
G. Dissertori, \underline{C. Martin Perez}\footnote{now at Laboratoire Leprince-Ringuet, Ecole Polytechnique, Palaiseau, France},
F. Nessi-Tedaldi
\Instfoot{eth}{Institute for Particle Physics and Astrophysics, ETH Zurich, Switzerland}
\end{Authlist}
\maketitle
\begin{abstract}
Fast hadrons have been observed to cause a cumulative damage in Lead Tungstate and LYSO crystals. The underlying mechanism has been proven to be the creation of fission tracks, which act as scattering centres, thus reducing the light collection efficiency. For calorimetry applications in an environment where large, fast hadron fluences are anticipated, predictions about damage in crystals are of great importance for making an informed choice of technology.
In the study presented here, simulations using the FLUKA package have been performed on Lead Tungstate, LYSO and Cerium Fluoride, and their results have been compared with measurements. The agreement that is found between simulation results and experimental measurements allows to conclude that the damage amplitude in a given material can be predicted with a precision that is sufficient to anticipate the damage expected during detector operation.
\end{abstract}
\vspace{5.5cm}
\conference{\em To be submitted to Proceedings 14$^{th}$ Pisa meeting on Advanced Detectors, La Biodola, Isola d'Elba (Italy)\\May 27$^{th}$ to June $2^{nd}$, 2018\\ for publication in Nuclear Instruments and Methods in Physics Research}
\end{titlepage}
\section{Introduction}

Experimental evidence exists that high-energy hadrons cause a specific damage in some inorganic crystals. It has been observed in Lead Tungstate\cite{r-LTNIM}, that such a damage component is cumulative and exhibits a wavelength-dependence $\left(\lambda^{-4}\right)$ of the absorption coefficient induced by irradiation, $\mu_{IND}$, which is typical for Rayleigh-scattering (RS)\cite{r-RS}. The scattered light, as demonstrated with Lasers, is polarised, yet another feature of RS. The cumulative behaviour has also been observed in LYSO~\cite{r-LYSONIM} and in BGO~\cite{r-BGOFN} while it is absent in Cerium Fluoride~\cite{r-CEFNIM}. RS requires the presence of scatterers with dimensions ${\cal O}(\lambda / 10)$, which have been advocated~\cite{r-LTNIM} to be damage tracks left by highly-ionising fission fragments. Fission is expected to occur only in materials containing elements with $ Z > 71 $, because these can undergo fission\cite{r-FISS}, and this matches the observations mentioned above. A visual proof of the presence of fission tracks has been provided~\cite{r-FISSNIM}, and thus the underlying mechanism can be considered understood beyond any reasonable doubt.

The damage amplitude depends of course on the material composition and on the hadron exposure. For the scintillating crystals considered in this study, a vanishing hadron-specific damage has been observed in Cerium Fluoride, while the ratio of induced absorption coefficients between Lead Tungstate and LYSO, determined at the peak wavelength of their scintillation emission, amounts to 4.5~\cite{r-LYSONIM}. Since the damage playing a role is one affecting the bulk of the crystal, it should be possible to perform simulations to reproduce the observed amplitudes thus validating the simulations, to further use them in a predictive manner.

For fission fragments to leave tracks, extensive studies in the past have shown that a high ionisation density is required. From Fig. 2 in \cite{r-FLE} and Fig. 1 in \cite{r-LTNIM} one deduces a minimum energy loss $dE/dx = 1\times 10^5$ MeV$/$cm, which is herein the main criterion used in the simulations to consider track creation.

\section{Simulation studies of hadron damage}
We have set up a simulation aiming at reproducing the experimental conditions of the hadron irradiations described in the quoted publications, by using the FLUKA package~\cite{r-FLUKA}. The simulation of the crystal irradiations by protons, as well as
the crystal geometries and compositions, are reproducing exactly the experimental setup of
the existing measurements. For 700000 primary protons simulated, each particle produced in the materials has been followed and at every step its energy loss has been computed. Consecutive steps satisfying the energy loss requirement for track formation have been joined to form segments  of damage tracks, against which light scattering is assumed to occur.

\begin{figure}[bht]
\begin{center}
 \begin{tabular}[h]{ccc}
{\mbox{\includegraphics[width=51mm]{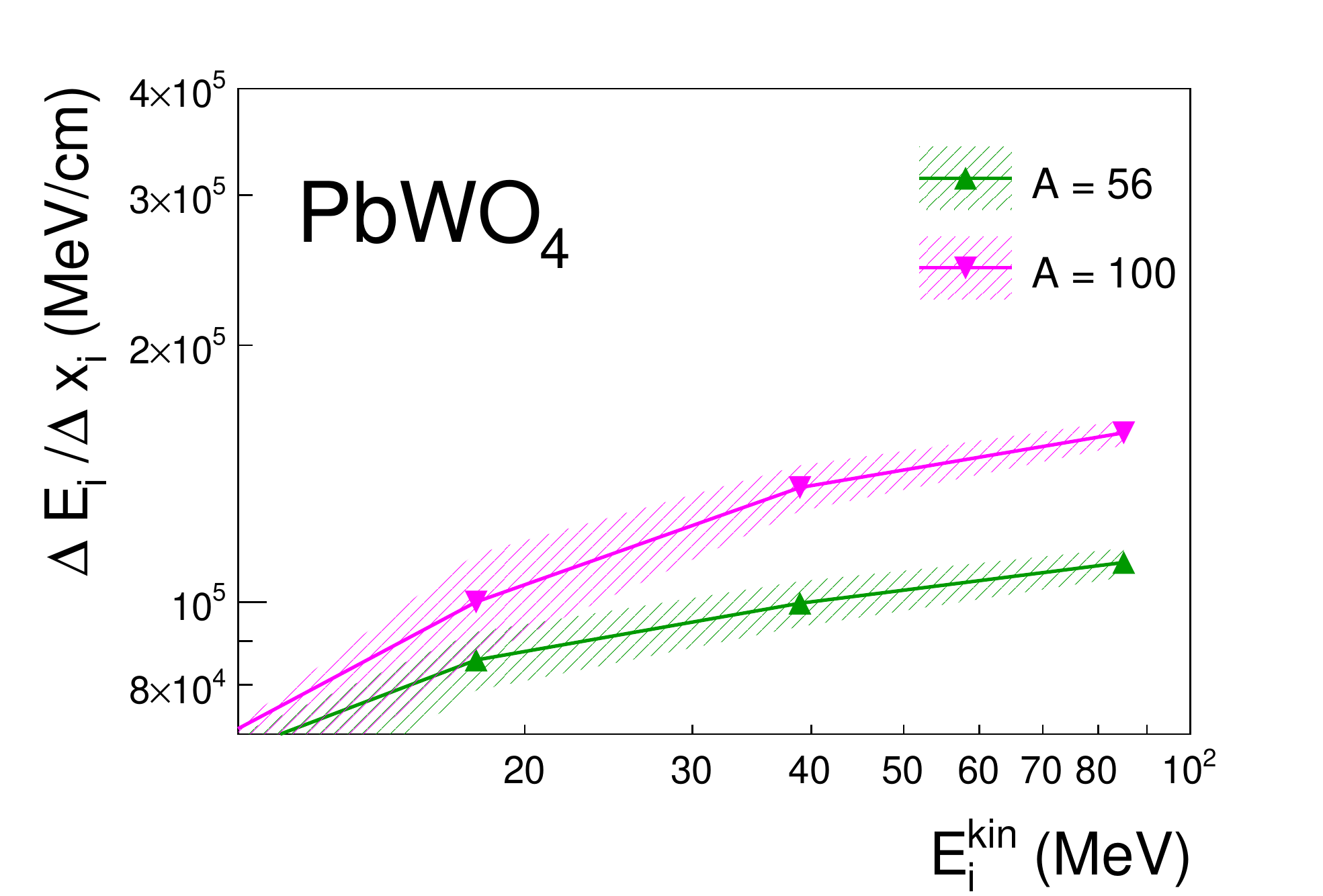}}} &
{\mbox{\includegraphics[width=51mm]{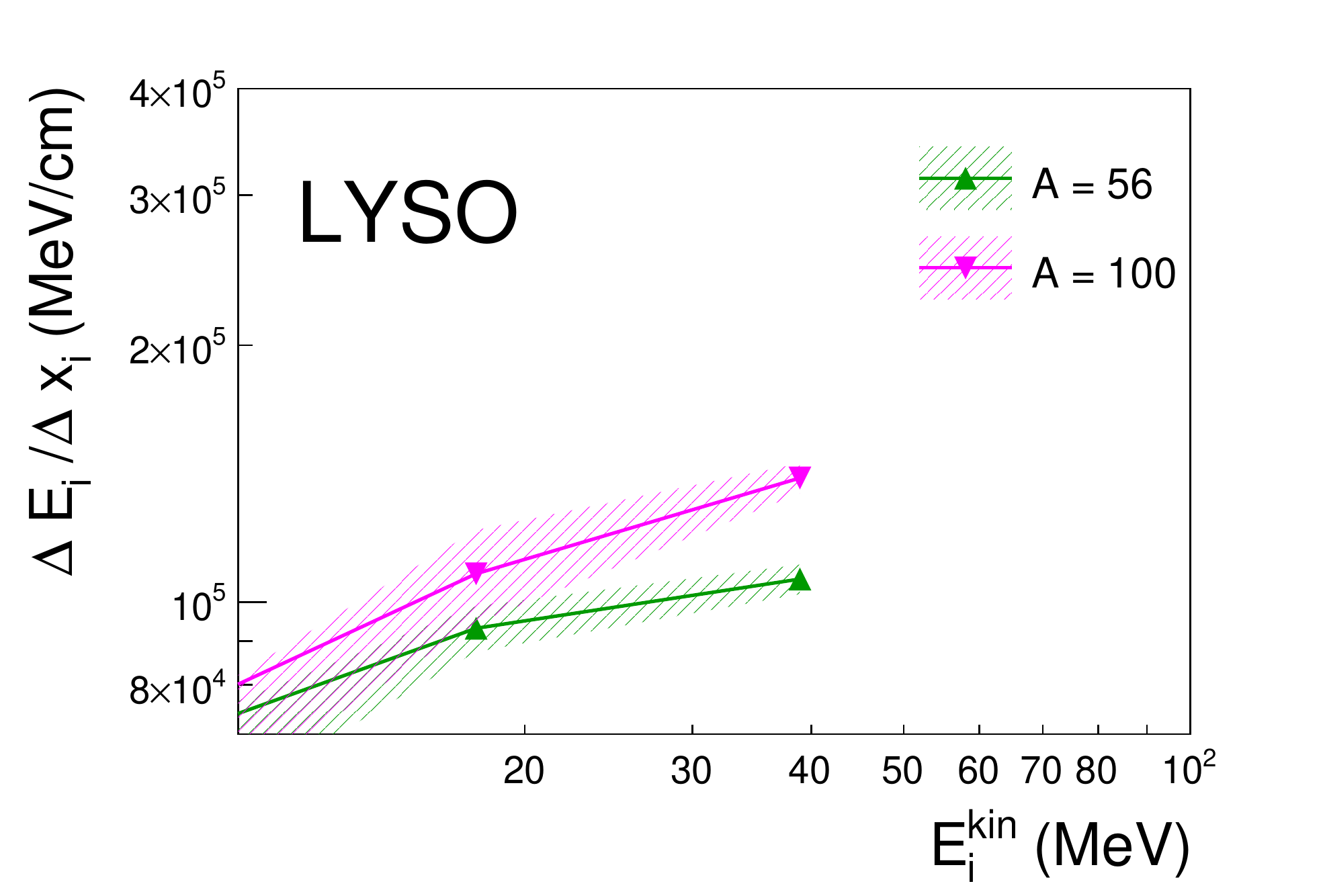}}} &
{\mbox{\includegraphics[width=51mm]{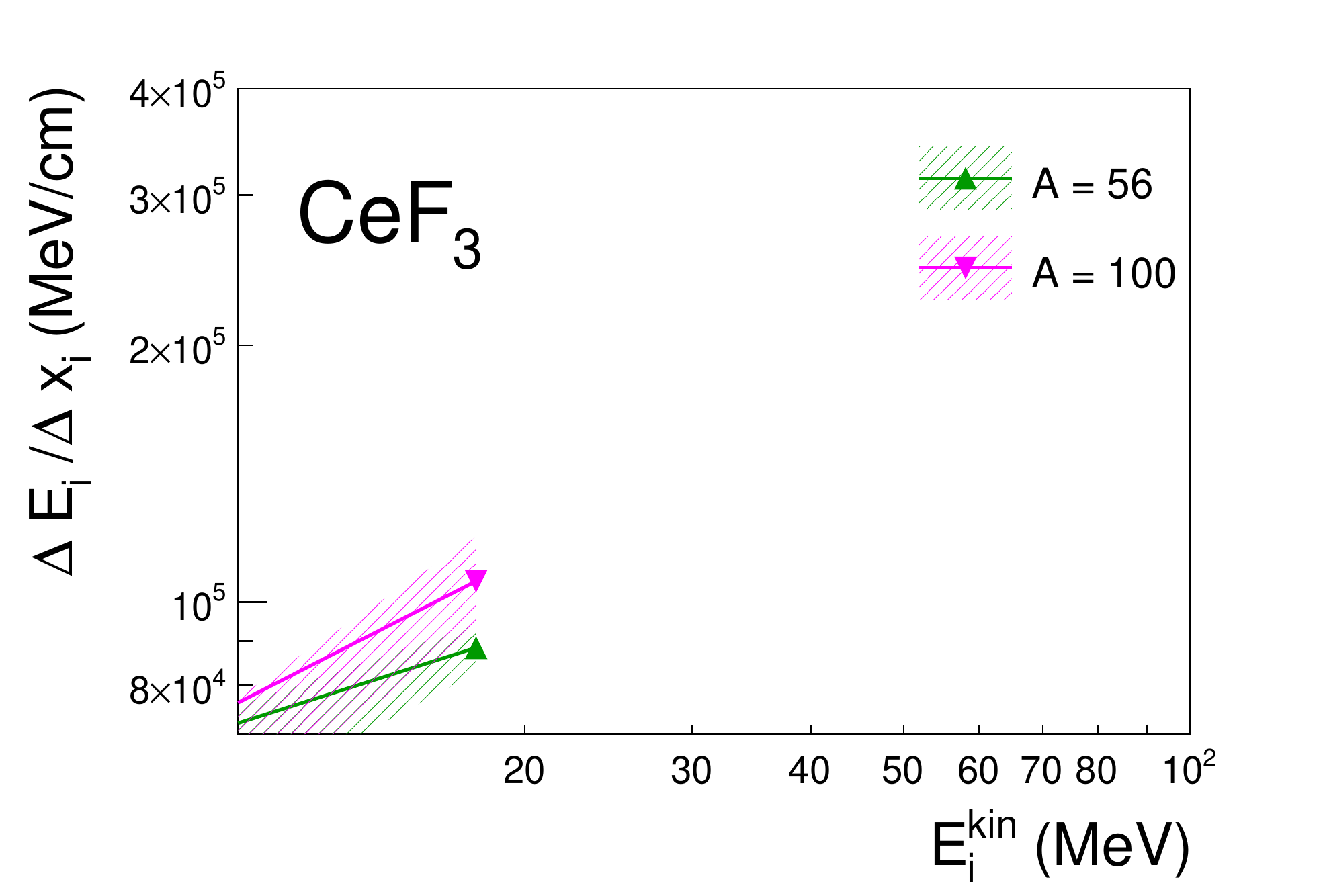}}}
   \end{tabular}
\end{center}
\caption{For Lead Tungstate (left), LYSO (center) and Cerium Fluoride (right), dE/dx values for different intervals of mass number A.}
\label{f-dEdx}
\end{figure}

Distributions of dE/dx values for each step are plotted as a function of kinetic energy in Fig.~\ref{f-dEdx}, grouped in bins of mass number A, for the highest values of energy loss observed. No entries appear for the lower bins centred around A = 6 and A = 30.
It is visible that in Cerium Fluoride hardly any fragments are produced that satisfy the requirement on energy loss for track creation. The result is validated by the lack of observed tracks in this crystal.
For Lead Tungstate and LYSO, only some of the highest A fragments reach the needed dE/dx, more prominently in the former, in agreement with experimental findings.

The FLUKA simulations yield a volume density of track segments $\rho_{seg} = 3.4 \times 10^{-3}$seg cm$^{-3}$/(p cm$^{-2}$) and a volume density of tracks $\rho_{trk} = 2.6 \times 10^{-3}$tracks cm$^{-3}$/(p cm$^{-2}$). This needs to be compared to experimental results~\cite{r-FISSNIM}, where the density of etched tracks crossing a surface was measured to be $\Phi = 1.8\times 10^{-6}$ tracks/p of an average length around $10 \mu$m. Agreement between the simulated volume density and the measured surface density is reached for an average track length $< L > = 7\mu$m, since this is the dimension that allows tracks in the volume to cross a surface, close to the observed value from the visualisation study~\cite{r-FISSNIM}, which yields an additional validation element for the simulations.

The Rayleigh Scattering cross section\cite{r-RS} is given by
\begin{equation}
\sigma_{RS}\; \alpha \; \frac{d^6}{\lambda^4}\left(\frac{n^2-1}{n^2+2}\right)^2
\end{equation}
with 
d the dimension of scatterers, $\lambda$ the peak-of-emission wavelengths for scintillation light (420 nm for PbWO$_4$ and 425 nm for LYSO), and 
n the index of refraction (2.2 for PbWO$_4$ and 1.82 for LYSO). The fraction of scattered light F is proportional to the number of scatterers, N$_{PWO}$ or respectively N$_{LYSO}$, but it can also be expressed in terms of $\mu_{IND}$, as $F = \mu_{IND}$L, with L the crystal length\cite{r-LTNIM}. From the ratio of scattered light in the two crystal materials, one obtains the relationship between induced absorption coefficients that were measured\cite{r-LYSONIM} and the simulated quantities:
\begin{equation}
R_{\mu} = \frac{\mu^{PWO}}{\mu^{LYSO}}= 1.8\times \frac{N^{PWO}}{N^{LYSO}}\times\left(\frac{d^{PWO}}{d^{LYSO}}\right)^6 
\end{equation}
Only scatterers with dimensions below $~\lambda/10$ participate in Rayleigh scattering. Since the distribution of observed segments length spans a few orders of magnitude, we have determined the simulated ratio as a function of the maximum allowed length d$_{max}$, and it was found to be $R_{\mu} = 2.5 \pm 1.0$,
where the uncertainty takes into account contributions from applying a common d$_{max}$ value and a common dE/dx threshold for PWO and LYSO, as well as from the uncertainties on the length ratio and on the value of the dE/dx ratio.

\section{Conclusions}
\label{s-CON}
FLUKA simulations of long crystals irradiated by 24 GeV protons producing hadron showers yield  no heavy, highly ionising fragments in CeF$_3$, as would be needed for track creation, in agreement with the absence of hadron specific damage. They yield, instead, heavy, highly ionising fragments in Lead Tungstate and LYSO, as needed for track creation, in agreement with the observed hadron specific damage.
The simulated  track densities in PWO are in agreement with experimentally observed ones, and finally, the simulated Rayleigh Scattering amplitude ratio between Lead Tungstate and LYSO is consistent, within uncertainties, with the measured one. Through all this corroborating evidence one can confidently use FLUKA simulations to estimate the order of magnitude of damage amplitude to be expected from hadrons in inorganic crystals.

\end{document}